# Intermittent Jolts of Galactic UV Radiation: Mutagenetic Effects

by
John Scalo, J. Craig Wheeler, and Peter Williams
University of Texas, Austin, Texas, USA

## ABSTRACT

We estimate the frequency of intermittent hypermutation events and disruptions of planetary/satellite photochemistry due to ultraviolet radiation from core collapse supernova explosions. Calculations are presented for planetary systems in the local Milky Way, including the important moderating effects of vertical Galactic structure and UV absorption by interstellar dust. The events are particularly frequent for satellites of giant gas planets at $\gtrsim$ 5-10 AU distance from solar-type parent stars, or in the conventional habitable zones for planets orbiting spectral type K and M parent stars, with rates of significant jolts about $10^3 - 10^4$ per Gyr. The steep source spectra and existing data on UVA and longer-wavelength radiation damage in terrestrial organisms suggest that the mutational effects may operate even on planets with ozone shields. We argue that the mutation doubling dose for UV radiation should be much smaller than the mean lethal dose, using terrestrial prokaryotic organisms as our model, and that jolts may lead to important real-time evolutionary episodes if the jolt durations are longer than about a week, corresponding to several hundred generation times, or much less if the equivalent of mutator genes exist in extraterrestrial organisms. Longer-term phylogenetic effects are likely if atmospheric photochemical disturbances lead to niche creation or destruction in relevant habitats.

## I. Introduction

Genetic diversity provided by mutagenesis is the raw material for natural selection and evolution. A significant fraction of current-day mutation is due to error-prone, light-mediated DNA damage repair of cyclobutane pyrimidine dimers induced by ultraviolet radiation (e.g. Alpen 1998, Jagger 1985). The near-universality of specialized mechanisms for DNA repair, including repair of specifically radiation-induced damage, from prokaryotes to humans (see Nickoloff & Hoekstra 1998a, b), suggests that the Earth has always been subject to damage/repair events above the rate of intrinsic replication errors. The antiquity of several of prokaryotes possessing a variety of radiation repair pathways suggests that the mechanisms developed very early in the development of life. It is clear that early organisms were subjected to significant ultraviolet radiation because anaerobic bacteria show intrinsic resistance to UV damage and use photoreactivation DNA repair of UV damage (Rambler & Margulis 1980), showing that most life did not evolve in the deep oceans. Any shielding or other UV-avoidance tactic, whether in surface waters (see Sagan 1973, Cleaves & Miller 1998) or on the Earth's surface (see Garcia-Pichel 1998) must have only served to moderate, but not obliterate, the solar UV flux. Additional evidence for early exposure to sunlight can be found in Brock et al. (1999) and in Xiong et al. (2000).

Lateral gene transfer is now believed to be a primary source of genetic diversity, and hence evolutionary driving force, in eubacteria and archaea (see Ochman et al. 2000, Doolittle 1999). The mechanisms involved in lateral gene transfer are often the same as those involved in the repair of DNA damage due to UV and ionizing radiation, suggesting a connection. Similar processes are involved in meiosis. These considerations open the possibility that radiation may have been the dominant generator of genetic diversity in the terrestrial past and on any extraterrestrial and extrasolar planets and satellites harboring life based on a genetic code. An excellent general review of the role of UV radiation in the development of early life can be found in Rothschild (1999)

We argue here that intermittent cosmic radiation can affect evolution through this variety of radiation-active mechanisms, restricting the present discussion to ultraviolet (UV) radiation. We also argue that such radiation can affect evolution indirectly through the creation and destruction of niches during intermittent disturbances of planetary photochemistry.

Sagan (1973) was among the first to discuss the potentially far-reaching effects of UV radiation on the first life forms. Even earlier, Sagan & Shklovskii (1966) suggested that supernova explosions could intermittently affect biological activity on the Earth. They pointed out that the UV and especially cosmic ray fluxes from supernovae could have been catastrophic for some organisms, but could have also been "favorable for evolution." Since then, a number of authors have suggested terrestrial biological effects due to supernovae and related phenomena (see Rudermann 1974, Crutzen and Bruhl 1996, Collar 1996, Dar, Laor & Shaviv 1996, Ellis & Schram 1995, Ellis, Fields & Schramm 1996). Nearly all this work focused on the possible relation of Galactic events to mass extinctions. Little attempt was made to estimate the rates at which significant events should have occurred as a function of flux or fluence, their potential for mutagenesis, or the significance of such events for life on other planets, satellites of giant planets, or extrasolar planets.

In the present paper, we concentrate on the 200-300 nm spectral region because DNA action spectra for many types of alterations peak at 260 nm and decline rapidly at larger and smaller wavelengths. Shorter wavelengths are relevant to atmospheric chemistry, as discussed below. Many cosmic sources have spectra that rise rapidly at longer wavelengths. The increased flux at larger wavelengths may compensate for the reduction in magnitude of biological sensitivity at these wavelengths, making the UVB, UVA (320-400 nm), and even the blue region of the spectrum potentially rich in biological affect. If so, then even planets with no ozone shielding will be susceptible to potential mutagenesis by UV events.

## II. Core-collapse Supernovae

Here we report results for one type of event: UV radiation from core-collapse (Type II) supernovae (SNe). Core collapse SNe produce UV radiation by two separate phenomena. First, there is a prompt hard UV radiation burst due to shock breakout, which is found in all numerical and analytical theoretical models (e.g. Falk 1978, Ensman & Burrows 1992, Matzner & McKee 1999). We estimate that such events typically produce about $10^{47}$ erg in hard UV photons, predominantly shortward of 200 nm, for about a day or less. The subsequent UV emission as the explosion progresses to the more protracted light curve phase, lasting 2-3 months while the outer hydrogen envelope expands and becomes transparent, has been computed only roughly. We estimate a UV energy release of again about $10^{47}$ erg or larger for the light curve phase, from both theory and observation, but with a luminosity of about $10^{41}$ erg $s^{-1}$, compared to the shock breakout luminosity of a few times $10^{44}$ erg $s^{-1}$. Thus both phenomena may give similar total UV energies (and hence fluences at a given distance from a particular planetary/satellite system), but with very different spectra, luminosities (and hence fluxes), and timescales. There are uncertainties in these numbers, but they are small (less than a factor of 10) compared to the uncertainties in the mutation doubling dose (§IIIA).

The issue of luminosities and timescales is crucial. Both shock breakout and light curve UV will be equally capable of affecting atmospheric chemistry (apart from differences in spectral distribution compared to photolysis cross sections). The smaller fluxes of the light curve UV compared to the shock breakout UV means that the former will be less capable of dominating the UV flux of the parent star. As discussed below, the duration of the explosion is also crucial with regard to evolutionary effects.

### III. Frequency of Significant Events

### A. Critical UV Fluence for Mutational Evolution

For high-energy (ionizing) photons or particles, the major mutational lesions are also those leading to lethality (e.g. double-strand breaks). The situation is very different for UV mutagenesis, a fact that has apparently been overlooked in the astrobiological literature. For $\lambda \lesssim 310$ nm (UVC and UVB), bipyrimidic photoproducts, most importantly cylclobutane pyrimidine dimers, are the dominant contributors to UV mutation (see Friedberg et al. 1995 for a comprehensive discussion; also Chandrasekhar & Van Houten 2000). These photoproducts dominate the mutation rate because they are repaired by error-prone photoreactivation and specifically bypassed by the SOS response, but are rarely lethal, thanks to accurate nucleotide excision and other repair pathways, leading to a potentially very small mutation doubling dose. The best-studied prokaryotic systems (*E. coli* and *B. subtilis*) exhibit significant lethality at fluences around $10^4$ erg $cm^{-2}$ (e.g. Dose et al. 1997), although significant variations occur, both smaller (e.g. Asad et al. 2000) and larger (e.g. spores, Dose et al. 1996; the likely Archaean thermophile *Chloroflexus aurantiacus* under anoxic conditions, Pierson et al. 1993; and the extreme radiation resistance of *D. radiodurans*, see Battista 1997). We can estimate a lower limit to the UV mutation doubling dose (MDD) for *E. coli* bacteria as follows. Drake (1991) found that among lower organisms the *spontaneous* mutation rate per genome is surprisingly constant (compared to the mutation rate per base pair, which varies by orders of magnitude). We therefore define the MDD (erg $cm^{-2}$) as the spontaneous mutation rate per genome, $SMR_G$, divided by the number of mutations per organism per unit fluence (erg cm$^{-2}$), multiplied by the number of germline genome copies per organism, which is one for prokaryotes. Drake et al. (1998) estimate a spontaneous mutation rate per base pair per generation for *E. coli* lac I as about 5.4 x 10$^{-10}$. The genome size is 4.6 x 10$^6$ bp, yielding $SMR_G \approx 2.4 \times 10^{-3}$. According to Hamkalo (1972), the dose at 254 nm to produce an average of one thymine dimer per *E. coli* DNA is $\approx 0.1$ erg cm$^{-2}$, so the number of thymine dimers produced per (erg cm$^{-2}$) is $\approx 10$. This gives an MDD of only $\sim 2 \times 10^{-4}$ erg cm$^{-2}$. This is a lower limit, since not all thymine dimers lead to mutation. Correction for cyclobutane dimers repaired by nucleotide excision repair (NER) and other essentially error-free repair pathways will be discussed in detail elsewhere; we estimate an order of magnitude correction factor of 10$^2$, giving MDD$\sim 0.02$ erg cm$^-2$ if we ignore timescale constraints on these repair paths. Our small UV MDD is also consistent with the fact that a single chain break or cross link is 2-3 orders of magnitude more likely to cause lethality than a single thymine dimer (Rahn 1972). Another way to see this is that the number of lesions per cell per lethal D$_{37}$ dose is 4 x 10$^5$ for thymine dimers in *E. coli*, but only about 100 for single strand breaks (Ward et al. 1997). These results strongly suggest that the mutation doubling dose for microorganisms due to UV radiation may be much smaller than the lethal dose. This is in contrast to the MDD for ionizing radiation, where the MDD is typically 0.1-0.3 times the lethal dose. The low MDD for UV combined with the likely influence of UVB, UVA, and blue light leads us to suspect that UV may be the dominant contributor to mutagenesis on extraterrestrial planets, whether or not a UVC screening agent exists.

Although we realize that extrapolation to other organisms (especially extraterrestrial ones!) is dangerous, these results strongly suggest that the mutation doubling dose for microorganisms due to UV radiation may be much smaller than the lethal dose. Given all the uncertainties, we adopted $F_{cr} = 600$ erg cm$^{-2}$ (closer to the lethality dose than to the estimated MDD) for both terrestrial and extraterrestrial microorganisms as a conservative fiducial estimate, with the understanding that this value is subject to order of magnitude uncertainties.

For the atmospheric attenuation of Archaen (pre-ozone) atmospheres, we adopt Cockell's (1998) calculations of absorption and scattering in $CO_2$ + $N_2$ atmospheres, giving an attenuation factor of about 3 in the 200 - 300 nm region. This is clearly a lower limit, since there may exist significant non-ozone UV shields, including aerosols. Adopting an attenuation factor of 3 means that the required fiducial fluence above the atmosphere is $F_{cr} = 2 \times 10^3$ erg cm$^{-2}$.

### B. Recurrence Rates

We have calculated the mean time between cosmic events that generate a fluence in excess of an estimated critical fluence (erg cm$^{-2}$) required for mutagenesis for a variety of astronomical phenomena. The *average* frequency of biologically and atmospherically significant events is surprisingly large. Assume that the Galactic events in question occur at a rate per unit volume, S. Then the frequency of events at a distance D is

$$f = (4\pi/3)SD^{-3}, \tag{1}$$

where D is the distance of the event. The mean time between events is T = 1/f. The distance from the event is related to the fluence F received in a given wavelength interval by $F = E/4\pi D^2$, where E is the total energy of the event in that wavelength interval. Solving for D and substituting in f above gives for the mean time between events at fluence level F:

$$T = (6\pi^{1/2}/S)(F/E)^{3/2} = 3.2 \times 10^{56} S^{-1} [F/E]^{3/2} \text{ yr}, \qquad (2)$$

where the event rate S is expressed in units of $yr^{-1}$ $pc^{-3}$ and F and E are in cgs units. Taking an average Galactic Type II supernova rate $S=1.5 \times 10^{-13}$ $yr^{-1}$ $pc^{-3}$ (Capellaro et al. 1997) and a total 200-300 nm energy release of $10^{47}$ erg as discussed above, $T_{SN} = 0.067$ $F_{cr}^{3/2}$ yr, and $F_{cr}$ is the critical fluence.

The order of magnitude formula for the mean interevent time, Eq. 2, is a severe underestimate at small fluences (or fluxes) because it neglects the density gradient perpendicular to the Galactic plane and the effect of interstellar dust extinction. Both of these effects eliminate a large number of very distant SNe. To correct for this, we assumed an exponential vertical distribution of Galactic dust and of SNe progenitors, with a scale height of 60 pc (see Binney & Merrifield 1988). For dust extinction we adopt 3 mag per kpc path length in the 200-300 nm spectral region (Binney & Merrifield 1988) with a scaling to the 200-300 nm spectral region based on Table 21.6 in Cox (2000). We neglect the radial density gradient of the Galaxy because its scale length (∼5 kpc) is so much larger than the scale lengths for vertical thinning and extinction.

The results are shown in Fig.1, where the mean time between events of given or larger UV fluence is shown. The short-dashed and long-dashed lines show the effects of vertical structure and dust extinction separately, while the solid line represents the combined effects. The dotted line corresponds to the homogeneous unattenuated approximation (eq.2). Notice that this plot is independent of any assumption about the planetary/satellite atmosphere (fluences are above-atmosphere), location in the local Galaxy, type of parent star, or selection of critical fluence. The effects of vertical structure and extinction reduce the mean recurrence time by only a factor of two or so for very large received fluences (above the atmosphere) of $10^5$ erg $cm^{-2}$, which nevertheless should occur roughly every 5 million years. We conclude that *all planets in any planetary system have been subjected to extremely large UV fluences (far above the lethality limit) about a thousand times.* For the smaller fluences required for mutagenesis in the example we have used for an Archaean atmosphere (2000 erg $cm^{-2}$) the recurrence time is increased drastically by the effects of Galactic structure and dust extinction relative to the simple estimate, but is still small, about $2 \times 10^5$ yr. This corresponds to about *20,000 such events above the 2000 erg $cm^{-2}$ level during the history of the solar system.* Considering that we may have underestimated the mutagenic sensitivity of even present-day micro-organisms because of the mutation/lethality disparity discussed earlier, this number could be much larger. As we show next, not all of these events are relevant, depending on the distance of the planet from, and spectral type of, the parent star, whose flux must be exceeded by the Galactic event if there is to be any significant effect.

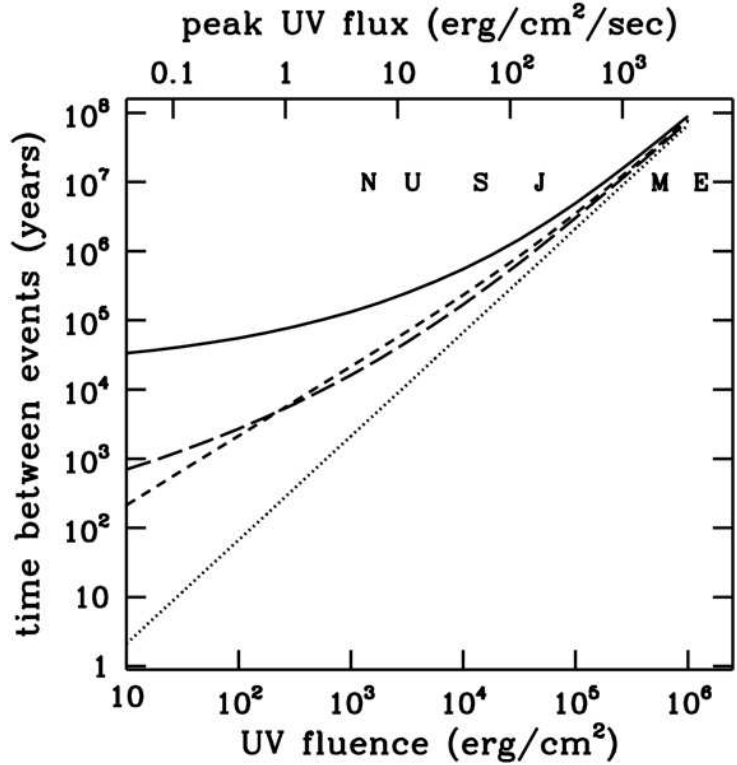

Fig. 1.— Time between receptions of a fluence in the 200-300 nm spectral region equal to or greater than that shown, for any position in the solar neighborhood, due to shock breakout and light curve events from core collapse supernovae. Long-dashed line: only effect of vertical Galactic structure included; short-dashed line, only effect of Galactic dust extinction included; solid line: both vertical structure and extinction included; dotted line: homogeneous no-extinction approximation (eqn. 2). The top horizontal axis shows a similar result, but for bolometric fluxes due to shock breakout events in core collapse supernovae. The inset letters show where to read off the present-day solar 200-300 nm fluxes at the position of each of the planets indicated, using the top horizontal scale. For atmospheric photochemistry the relevant wavelengths are smaller than for biological affect, and the planetary fluxes should be reduced by a factor of about five (inset letters shifted to left by this factor).

### IV. Fluxes and Atmospheric Chemistry

The Galactic UV events are only relevant if the flux from the Galactic event exceeds the background parent star flux. Shown on the upper horizontal axis of Fig. 1 are the received bolometric fluxes for shock breakout from SNe ($\sim 10^{44}$ erg s$^{-1}$) while the inset indicates the flux from the present-day Sun at the distances of the various planets. Events exceeding the solar flux by a significant factor should have occurred several times even during the history of the Earth, but only occur frequently (more than thousands of times) for Jupiter and beyond. Alternatively, for extrasolar planets orbiting K and M spectral type stars, the effects are frequent even in the traditional habitable zone, because the fraction of the photospheric flux in the UV should be smaller by roughly an order roughly of magnitude for these stars (see Kasting 1997, Cockell 1999), and the recurrence time involves this flux to the 3/2 power.

UV jolts exceeding the UV flux of the Sun that are important for photochemistry should occur much more often than those important for mutagenesis. If the background solar UV flux at wavelengths below the photolysis thresholds of most photochemically important molecules is taken to be $10^3$ erg cm$^{-2}$ s$^{-1}$ at 1 AU (the thresholds are almost all in the 100-200 nm range, see for example Griffith et al. 1998), then, for a planet or satellite at a distance R$_{AU}$ from the Sun (in units of AU), a random SN breakout will exceed this flux with an average time interval T$_{SN}$ $\sim 10^7/\text{R}_{AU}{}^3$, adopting L = $10^{11}$ L$_\odot$ for the average UV breakout event. More detailed calculations including the effects of vertical Galactic structure and dust extinction are shown in Fig.1 (fluxes given on top horizontal axis). The corresponding positions of the Solar flux at the distances of the planets should be shifted to the left by a factor of five or so to account for the smaller Solar flux below the photolysis thresholds relative to the 200-300 nm flux used in the plot.

This result indicates that atmospheric photochemistry must have been significantly perturbed about $10^2$ - $10^3$ times during the history of the solar system, even at 1AU. For the Jovian planets and their satellites, the number of such jolts is much larger, by a factor of order $10^3$, because of the lower solar background at larger heliocentric distances. In the latter case, photochemically significant jolts ($\lambda \lesssim 100 - 200$ nm) must occur roughly every $10^5$ yr! *Photochemistry of the outer planets and their satellites has likely been disturbed at least 50,000 times during the history of the solar system.* The frequency would again be roughly $10^{3/2}$ times larger for cooler parent stars than the sun because of the order of magnitude decrease in UV flux.

### V. Evolutionary Change?

In order for a transient mutagenic event to result in evolutionary consequences, the mutation must not only be passed on to a subsequent generation on an individual basis, it must also persist for a sufficiently large number of generations that fixation occurs; all members of the population must come to possess that mutation. Only if the mutation fixates *and* is adaptive in terms of the stress caused by the enhanced radiation flux (or the environmental changes wrought by the radiation) will the episodic cosmic events have evolutionary significance. The duration of peak UV flux from a core collapse SN in the 200-300nm spectral region is estimated to be from a day to a week. A week corresponds to several hundred bacterial regeneration times, assuming a typical generation time of $10^3$ sec. For comparison, a similar statement for humans would imply that evolutionary change is possible over about 2 x$10^4$ yr.

Are such rapid fixations possible? Microorganisms apparently possess great facility for rapid evolution compared to multicelled macroscopic organisms. Experimental evidence for rapid evolutionary bursts in microbial populations is now abundant (e.g. Elena, Cooper, & Lenski 1996, Coyne & Charlesworth 1996). Most relevant to the present paper are the experiments by Ewing

– 8 –

(1995, 1997) in which radiation resistance was increased in the presence of increasingly stronger X-ray and UV doses, and with timescales that correspond to hundreds of generations. Although the time between exposures was many orders of magnitude smaller than would occur for cosmic events, it is still significant that the total exposure time for appreciable adaptation was only about 10 hr. Obviously the question is complex, but there is probably sufficient evidence to argue for widespread mutation fixation due to Galactic radiation events with durations in excess of a week. For this reason, the enhanced cosmic ray flux due to the (later) passage of the supernova remnant near the planetary system in question may play a more important role, since the duration could be many centuries or longer.

Mutator genes, which act to increase the spontaneous mutation rate in response to environmental stresses, is an extremely active area of research (see Boe et al. 2000, Lieber 2000), especially because of its potential relevance to carcinogeneisis (Cairns 1998, 2000), and may be expected to play a significant role in the response to a sudden elevation of the UV flux. The idea that stress-inducible processes exist that operate only when high mutation rates are advantageous is supported by several recent identifications of mutator genes and their associated mutases (see Radman 1999, Masutani et al. 1999, Friedberg & Gerlach 1999). Among the most intriguing results is that of Sniegowski et al. (1987) who observed spontaneously arising mutator genes in *E. coli* adapting to a new environment. Several authors have suggested that adaptive evolution by mutator genes interacting with an unpredictable environment may be beneficial to bacterial cells and populations; the same speculation could be made for extraterrestrial microbial organisms if they are genetic code lifeforms.

There is also the possibility for "adaptive mutations" in which beneficial mutations occur within a single generation, i.e. during the resting state of a cell. The prevalence and mechanisms of this phenomenon are still controversial over a decade after their putative discovery (see Foster 1999 for a review). An optimistic theoretical interpretation of adaptive mutation is given by Koch (1993), who argued persuasively that bacteria have evolved in such severely fluctuating environments that they have evolved a "catastrophe kit" to deal with sudden environmental changes. The components of such a kit include the development of metabolically inactive states (e.g. sporulation), activation of previously evolved but silent genes, increasing rates of mutation under stress, and activation of exogenous gene transfer (transformation, conjugation, plasmid transmission). We suggest that frequent intermittent blasts of radiation from cosmic sources may have played a pivotal role in the development of this toolkit. If adaptive mutation is possible in an enhanced radiation environment, then the timescale argument discussed above is irrelevant, and the evolutionary consequences of Galactic radiation sources become much more likely, even for events of very small duration, as long as the frequency of high-flux and high-fluence doses is large enough.

We acknowledge extremely useful conversations and correspondence with Charles Cockell, Andrew Karam and Peter Höflich. This work was supported by NSF Grant 9907582.